\documentclass[amssymb,useAMS,prd,aps,amsmath,nofootinbib,superscriptaddress,twocolumn]{revtex4}

\usepackage{natbib}
\usepackage{color}
\usepackage{graphicx}
\usepackage{amsmath}
\usepackage[caption=false]{subfig}
\usepackage[colorlinks=true,linkcolor=blue,citecolor=blue,urlcolor=black]{hyperref}

\begin{document}
\title{Glow in the Dark Matter: Observing galactic halos with scattered light}
\author{Jonathan H. Davis}
\affiliation{Institut d'Astrophysique de Paris, 98 bis boulevard Arago, 75014 Paris, France}
\affiliation{Institute for Particle Physics Phenomenology, Durham University, Durham, DH1 3LE, UK}
\author{Joseph Silk}
\affiliation{Institut d'Astrophysique de Paris, 98 bis boulevard Arago, 75014 Paris, France}
\affiliation{Department of  Physics \& Astronomy, The Johns Hopkins University \\
3400 N Charles Street, Baltimore, MD 21218, USA}
\affiliation{Beecroft Institute of Particle Astrophysics and Cosmology, Department of Physics,
University of Oxford, Denys Wilkinson Building, 1 Keble Road, Oxford, OX1 3RH, UK
 \\ {\smallskip \tt  \href{mailto:jonathan.h.m.davis@gmail.com}{jonathan.h.m.davis@gmail.com}, \href{mailto:silk@iap.fr}{silk@iap.fr}\smallskip}}

\begin{abstract}
We consider the observation of diffuse halos of light around the discs of spiral galaxies, as a probe of the interaction cross section between Dark Matter and photons. Using the galaxy M101 as an example, we show that for a scattering cross section at the level of $10^{-23} \cdot (m / \mathrm{GeV})$~cm$^2$ or greater Dark Matter in the halo will scatter light out from the more luminous centre of the disc to larger radii, contributing to an effective increased surface brightness at the edges of the observed area on the sky. This allows us to set an upper limit on the DM-photon cross section using data from the Dragonfly instrument. We then show how to improve this constraint, and the potential for discovery, by combining the radial profile of DM-photon scattering with measurements at multiple wavelengths. Observation of diffuse light presents a new and potentially powerful way to probe the interactions of Dark Matter with photons, which is complimentary to existing searches.
\end{abstract}

\maketitle

\section{Introduction}
Data from rotation curve surveys indicate that the kinematical behaviour of luminous matter, such as stars, in galaxies can not be explained purely by their own gravitation~\cite{Klypin:2001xu,2013A&A...557A.131M}. This implies either that our theory of gravity is incorrect at large scales, or that there is an additional component of matter in our galaxy which we have yet to observe. Indeed the latter scenario is compelling based on several additonal observations (e.g. the Cosmic Microwave Background (CMB) measured most recently by Planck~\cite{Ade:2013zuv} and lensing in clusters~\cite{Newman:2012nw}), and such rotation curves can be explained by the presence of a massive and approximately spherical halo of Dark Matter (DM) surrounding the galactic disc.

Since these halos of DM have not been directly observed (apart from via their gravitation) it is reasonable to assume that they are composed of electrically neutral particles, which by definition do not scatter light. However this is not strictly necessary. Indeed many models of Dark Matter include a small coupling with photons~\cite{Foot:2004wz,Cline:2012is}, and DM-photon interactions have been previously discussed in terms of their effect on the CMB~\cite{Wilkinson:2013kia,Cyr-Racine:2013fsa,CMBcharge,McDermott:2010pa}, the shape of elliptical galaxies (via DM self-interactions)~\cite{McDermott:2010pa,Peter:2012jh} and large-scale structure~\cite{Boehm:2014vja,Wilkinson:2013kia,Buckley:2014hja}. However there are few direct constraints on the DM-photon cross section, especially as a function of photon wavelength.
Furthermore some Dark Matter candidates (e.g. axions and axion-like particles~\cite{Jaeckel:2013uva} or asymmetric dark matter~\cite{Kaplan:2009ag}) may only show up through their couplings to photons, while evading more traditional search strategies such as Direct Detection~\cite{LUX,Aprile:2012_225,Agnese:2014aze}, Indirect Detection~\cite{Daylan:2014rsa,Gordon:2013vta,Boehm:2014hva,Abdo:2010ex} or colliders~\cite{Malik:2014ggr}.

Hence in this work we consider the prospect for observing DM halos in spiral galaxies directly through the light they scatter from the disc, if the DM-photon interaction cross section is non-zero.
We have two main aims: firstly we seek to quantify to what extent the DM can scatter light from the disc of a spiral galaxy and still remain unobserved, and secondly we examine prospects for observing DM-photon scattering in the future. 
In section~\ref{sec:m101} we show, using the galaxy M101 as an example, that particles of DM can scatter light from the more luminous centre of the disc out to the edge, where the scattered signal is potentially competitive with the emission from the disc itself (for a large enough scattering cross section). This light is very faint, however we show that the Dragonfly instrument~\cite{2014ApJ...782L..24V} is sensitive enough to detect this signal for Thomson-like cross sections.

In section~\ref{sec:spectra} we show that the spectrum of light as a function of wavelength can be used to separate our signal from potential backgrounds, such as dust-scattering or a halo of older stars.
In section~\ref{sec:lims} we discuss uncertainties in our knowledge of the backgrounds to a dedicated search, in section~\ref{sec:comp} we compare the sensitivity of our method to previous constraints from the CMB and large-scale structure and we conclude in section~\ref{sec:conc}.

\section{Example constraints from M101 \label{sec:m101}}
\begin{figure}[t]
\includegraphics[scale=0.27]{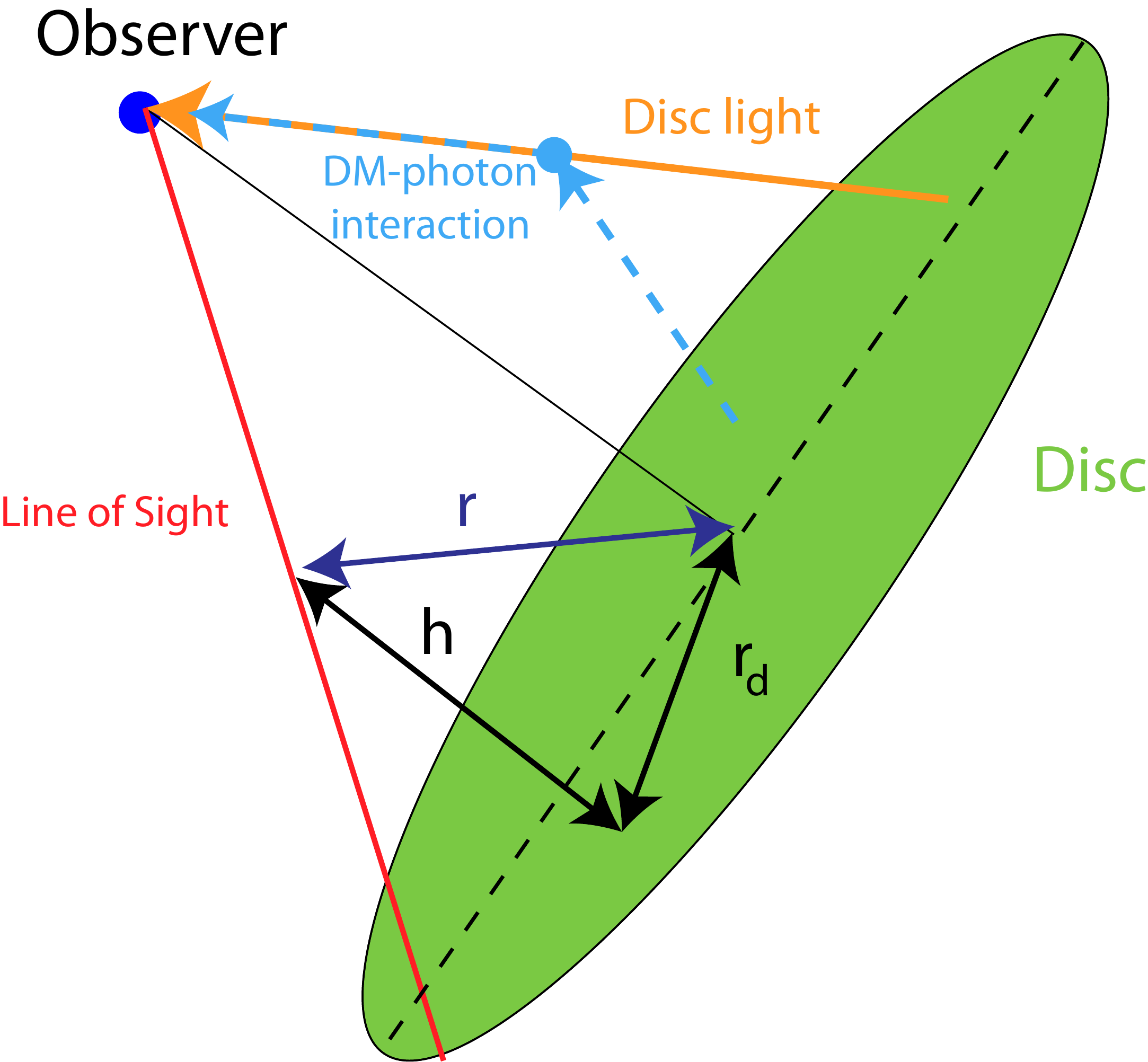}
\caption{A light ray from the inner parts of the disc, where the luminosity is larger, can scatter with a Dark Matter particle in the halo, thereby altering its path. Hence, for example, the dashed blue light ray will appear to originate from the outer parts of the disc. This will compete with light which does not scatter on its way to Earth, as shown by the orange arrow. There will also be emission from a stellar halo and scattering from dust outside of the disc, which we do not show here.}
\label{fig:scattering_diag}
\end{figure}

To illustrate our idea we consider the galaxy M101 as an example, for which observations of its surface brightness have been made using the Dragonfly instrument~\cite{2014ApJ...782L..24V}  (see also refs.~\cite{2014A&A...567A..97S,0004-637X-739-1-20} for other galaxies). We examine the possibility that some component of the light, observed to originate from a particular point on the disc of M101, could actually arise from photons emitted at another point on the disc which have been scattered by DM particles somewhere along the observer's line of sight. Hence the total luminosity of the disc is unaffected, but the radial profile of observed light is flattened at large radii.

This process is shown graphically in figure~\ref{fig:scattering_diag}. The orange solid arrow shows the path of a photon from the disc to the observer on Earth. This is accompanied by a photon from a more central region of the disc (dashed blue line), which has scattered with a DM particle somewhere in the halo, altering its direction to make it appear to originate from a position towards the edge of the disc instead.
This is somewhat analogous to scattering by dust~\cite{2014ApJ...789..131H}, which will also contribute to the apparent surface brightness, along with emission from the halo of older stars around the disc.

This scattering can occur anywhere along the observer's line of sight $l$. For an element $\mathrm{d} l$ of the line of sight distance the fraction of scattered photons from DM will be proportional to $n_{\mathrm{DM}} \sigma_{\mathrm{DM}-\gamma} \mathrm{d} l$, where $n_{\mathrm{DM}}$ is the number density of DM particles and $\sigma_{\mathrm{DM}-\gamma}$ is the scattering cross section (which may be a function of the frequency/wavelength of the scattered light). The total flux of scattered light along the line of sight is then the integral of this scattering fraction over all $l$. We assume that the cross section is small enough such that each photon scatters at most once with a DM particle, and additionally that the DM does not emit photons through e.g. decay or self-annihilation.

\begin{figure}[t]
\centering
\includegraphics[width=0.48\textwidth]{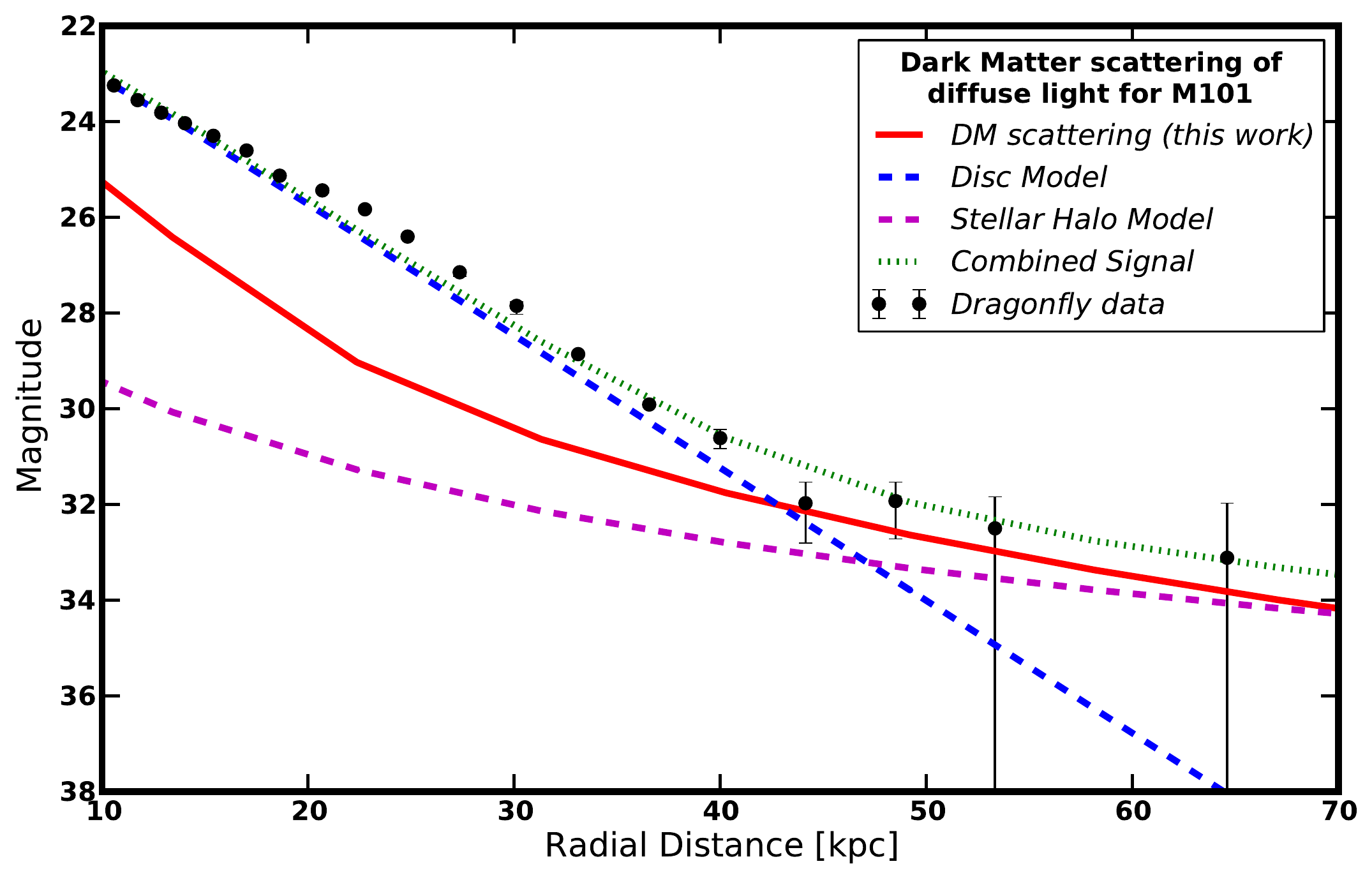}
\caption{DM-photon scattering profile for a cross section of $\sigma_{\mathrm{DM}-\gamma} = 10^{-23} (m_{\chi} / \mathrm{GeV})$ cm$^2$ compared to Dragonfly data for M101~\cite{2014ApJ...782L..24V}. Here `Magnitude' refers to the surface magnitude through an SDSS g-band filter.
The signal from Dark Matter - photon scattering is stronger than that from the disc at large radial distances, but can not easily be separated from the stellar halo emission. For the stellar halo fraction we use the best-fit value from ref.~\cite{2014ApJ...782L..24V}.}
\label{fig:dm_photon_-28}
\end{figure}

As such the total emission observed from a point on the sky (denoted by the angles $\theta_e$ and $\phi_e$) is the sum of three components: Dark Matter scattering, emission from the disc and emission/scattering from old stars and dust in the halo $\Phi_{\mathrm{halo}}$ respectively~\cite{0004-637X-712-1-692},
\begin{equation}
\begin{split}
\Phi (\theta_e,\phi_e) = \int \frac{\mathrm{d}l \, \sigma_{\mathrm{DM}-\gamma} n_{\mathrm{DM}}(r) \mathcal{L}_d(h,r_d)}{4 \pi l^2} \\ + \frac{\mathcal{L}_s(r_d(\theta_e,\phi_e))}{4 \pi l_d^2} + \Phi_{\mathrm{halo}},
\end{split}
\end{equation}
where we assume that the DM density $n_{\mathrm{DM}}$ depends only on the spherical radial distance from the galactic centre $r$ and $\mathcal{L}_s (r)$ is the luminosity of the disc surface per unit area.  
The function $\mathcal{L}_d (h,r_d)$ is the luminosity of light at a height $h$ from the disc and radial distance $r_d$ from its centre (see figure~\ref{fig:scattering_diag}), and is given by
\begin{equation}
\mathcal{L}_d (h,r_d) = \int \frac{ \mathrm{d} \alpha  \, \mathrm{d} \theta \, \alpha \mathcal{L}_s(\alpha)}{4 \pi (h^2 + \alpha^2 \sin^2 \theta + (\alpha \cos \theta - r_d)^2)}.
\end{equation}
There should also be an effective dimming of the emission from the disc due to the scattering from DM, which we assume to be negligible in this work.

We are now in a position to compare our predictions to observational data for M101 \cite{2014ApJ...782L..24V}. We use the exponential disc+bulge profile for $\mathcal{L}_s$ and we assume that $\Phi_{\mathrm{halo}}$ is dominated by emission from the stellar halo, for which we use the surface brightness profile given in \cite{2014ApJ...782L..24V}, rather than dust {(this should be true for optical wavelengths as dust scatters mainly ultra-violet light~\cite{2014ApJ...789..131H,Menard21062010}, a point we return to later on)}. For the Dark Matter density we use a Navarro-Frenk-White distribution (NFW)~\cite{Navarro:1996gj}. We take the distance to M101 to be $l_d = 7$~Mpc~\cite{2041-8205-760-1-L14}, and have made the assumption that the disc is completely face-on for simplicity. {If the disc were more edge-on its emission would be reduced at large radii, making the DM-photon signal easier to observe.}

The result is shown in figure~\ref{fig:dm_photon_-28}.
Since the profile of light from DM scattering depends on both $\mathcal{L}_s$ and the spherical distribution of the DM itself $n_{\mathrm{DM}}(r)$, it is much flatter than that from the disc. This therefore contributes to an \emph{apparent} brightening of the disc at large radii, as some of the light from the more luminous centre of the disc is scattered out to the edges, where the emission from the disc itself is lower. 

The situation in figure~\ref{fig:dm_photon_-28} is complicated by emission from the stellar halo, which can also contribute significantly at large radii~\cite{0004-637X-739-1-20,2014ApJ...782L..24V}. The magnitude of the stellar halo emission is not known \emph{a priori} for M101 and so for this data alone we are only able to set an upper bound on the scattering cross section $\sigma_{\mathrm{DM}-\gamma}$, by marginalising over the stellar halo fraction (we employ a flat prior for its amplitude). Our upper limit at $90\%$ confidence is at the level of $\sigma_{\mathrm{DM}-\gamma} \lesssim 10^{-23} \cdot (m / \mathrm{GeV})$~cm$^2$.
Since the Dragonfly measurements were made using g and r band filters, this constraint applies to wavelengths around $\lambda \sim 500$~nm. 
{This upper limit is conservative and robust to changes in the background model. Indeed the weakest possible limit is set when we assume no stellar halo background, such that the DM-photon scattering signal makes up \emph{all} of the observed light for $r \gtrsim 50$~kpc. In this case our upper limit is $\sigma_{\mathrm{DM}-\gamma} \lesssim 2 \cdot 10^{-23} \cdot (m / \mathrm{GeV})$~cm$^2$ i.e. only a factor of two weaker than when marginalising over the stellar halo contribution.}

\section{Using Spectral Information to Separate Dark Matter from Dust and the Stellar Halo \label{sec:spectra}}

We have shown that it is possible to obtain bounds on the DM-photon scattering cross section using measurements of diffuse light. 
However even though the disc emission is small at large radii, there are additional backgrounds such as emission from a halo of older stars, which make discovering DM-photon scattering difficult.
In this section we consider methods of overcoming these issues and improving the sensitivity to DM-photon scattering. 

\begin{figure}[t]
\centering
\includegraphics[width=0.49\textwidth]{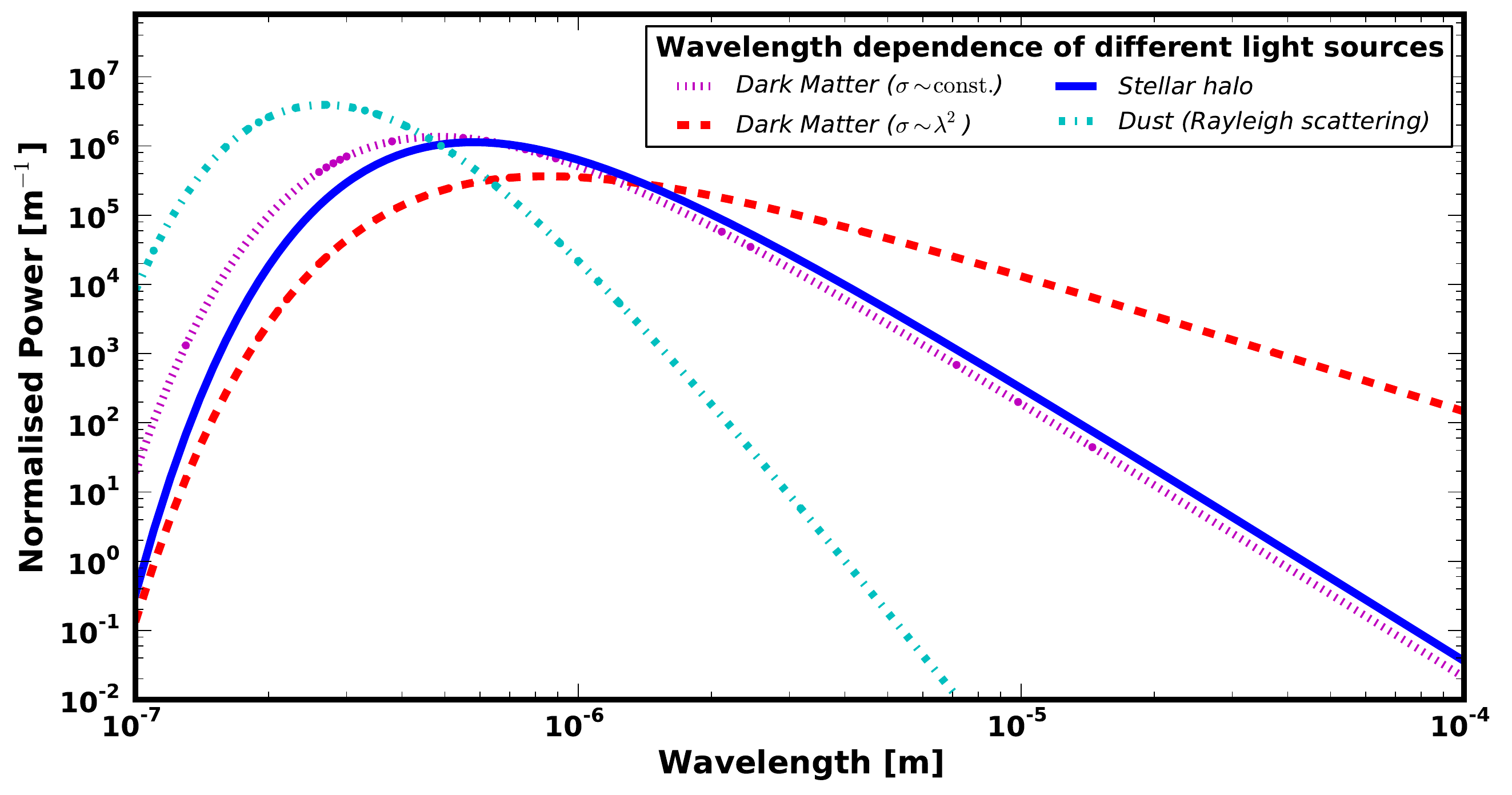}
\caption{Spectra of scattered light from Dark Matter as a function of wavelength $\lambda$, for both a constant cross section and one which varies as $\lambda^2$. This is compared with emission from the stellar halo and from Rayleigh scattering due to dust particles. At longer wavelengths DM with $\sigma_{\mathrm{DM}-\gamma} \sim \lambda^2$ should be easier to distinguish from potential backgrounds.}
\label{fig:dm_photon_spectra}
\end{figure}

There are broadly two ways of achieving increased sensitivity: taking measurements of surface brightness away from the disc, which would reduce emission from the disc itself, and using multi-wavelength information to separate sources based on their spectra. For the latter we illustrate this point in figure~\ref{fig:dm_photon_spectra}, showing the expected spectra as a function of wavelength $\lambda$ for different DM and background sources. 

We focus on two different types of background which should be present towards the outskirts or away from the disc: scattering from dust particles~\cite{2012MNRAS.423L.117M,Menard21062010,2014ApJ...789..131H} and emission from older stars in the halo~\cite{0004-637X-712-1-692} (there are other light sources present in the halo which emit at short wavelengths, such as hot gas~\cite{Li11102008} whose emission peaks in the X-ray band).
We have assumed a simplistic model for the dust, in which the dust particles are smaller than the wavelength of scattered light, such that the cross section is that for Rayleigh scattering $\sigma_R \sim \lambda^{-4}$. Hence the spectrum from dust-scattering peaks in the ultra-violet (UV) range~\cite{2014ApJ...789..131H,Menard21062010}. {Since the stellar halo is composed mainly of older stars~\cite{Cooper01082010,Cooper:2013gva,Foster21082014,Lowing:2014tza}} we have assumed that the spectrum is that of a typical red star i.e. a black body with a temperature of 5000K, while for the disc we assume a black body with a temperature of 6000K, similar to the Sun. {This assumption is approximate as both components are actually composed of stars with a range of colours. Indeed the disc likely possesses a colour gradient, with the younger/bluer stars present towards the outer edges~\cite{Bell01032000}.}

What is clear from figure~\ref{fig:dm_photon_spectra} is that the potential for observing DM-photon scattering above astrophysical backgrounds depends strongly on how the scattering cross section varies with $\lambda$. For example scattered light from Dark Matter with $\sigma_{\mathrm{DM}-\gamma} \sim \lambda^2$ will be more visible at longer wavelengths, where the contribution from other sources should be smaller (the same would be true for $\sigma_{\mathrm{DM}-\gamma} \sim \lambda^4$).  Such a dependence may arise for a DM-photon cross section which scales inversely with the square (or fourth power for $\sigma_{\mathrm{DM}-\gamma} \sim \lambda^4$ models) of the photon energy. {This could proceed through the exchange of a light mediator particle (e.g. ~\cite{Jaeckel:2013ija}), with the scale at which the cross section stops increasing with $\lambda$ depending on the mediator mass.}
Hence these models of DM can give observable signals even if their \emph{integrated} emission is less intense than that from e.g. dust or the stellar halo.

\section{Limitations due to uncertainties in the backgrounds \label{sec:lims}}
Dark Matter scattering of light from the disc has a potentially unique radial profile (figure~\ref{fig:dm_photon_-28}) and spectrum (figure~\ref{fig:dm_photon_spectra}) which is difficult to mimic with astrophysical sources. Hence it should be possible to improve the constraint on $\sigma_{\mathrm{DM}-\gamma}$ from section~\ref{sec:m101} if we were to combine information from the radial profile with measurements at multiple wavelengths into a statistical analysis of many galaxies, and in particular if observations can be made away from the disc. 

However the detectability of our DM-photon signal depends crucially on how well the disc and stellar halo components are understood at large radii. Indeed at present it is not clear how accurate the analytical models used for M101 are for other galaxies~\cite{Cooper01082010}. For example some galaxies posses significant streams towards their outskirts~\cite{MartinezDelgado:2008fh,Foster21082014} which complicate the radial and spectral profile of stars in the halo. In addition the exponential disc model can vary considerably between galaxies even at large radii e.g. it can be truncated past a certain radius or warped~\cite{Dutton:2008hh,Minchev:2012uy,2010MNRAS.402..713S,warped_discs}.

This problem would be mitigated by using observations of diffuse light around many galaxies, since e.g. although stellar streams~\cite{MartinezDelgado:2008fh,Foster21082014} may be able to mimic the DM spectral and radial profile for one galaxy, it is highly unlikely that this could occur at similar radii and with similar patterns for many different galaxies. 
Ideally, stacking a number of outer profiles with rescaled disk components would be an optimal approach for addressing these un- certainties. However we can only have confidence in a stacking approach once we have an improved statistical understanding of how the stellar halo and disc profiles and spectra vary from galaxy to galaxy.
Although recent progress has been made in for example simulating stellar halos~\cite{Cooper01082010,Cooper:2013gva,Foster21082014,Lowing:2014tza} such a statistical analysis is not presently possible.
 However since $\sigma_{\mathrm{DM}-\gamma}$ is poorly constrained over a wide range of frequencies only a basic knowledge of the backgrounds is needed to progress here.

\section{Comparison with existing constraints and searches \label{sec:comp}}
The DM-photon interaction cross section is relatively unconstrained directly. 
The strongest constraint comes from the damping of sub-halo scale structures by DM-photon interactions~\cite{Boehm:2014vja,Boehm:2004th}, leading to a $90 \%$ confidence limit of $\sigma_{\mathrm{DM}-\gamma} < 5.5 \cdot 10^{-9} \cdot (m / \mathrm{GeV}) \cdot \sigma_T$~\cite{Boehm:2014vja}, where $\sigma_T = 6.65 \cdot 10^{-24}$~cm$^2$ is the Thomson cross section.
As already discussed the size of the DM-photon cross section could be dependent on wavelength, hence the relative sensitivity of our method compared to that in ref.~\cite{Boehm:2014vja} depends on the photon spectrum at the time of DM-photon decoupling. If we take this to be the temperature of the CMB at recombination i.e. around 3000K then the photon population is that of a black body with maximum intensity around a wavelength of $10^{-6}$~m. Hence this constraint applies to $\sigma (\lambda \approx 10^{-6} \, \mathrm{m})$.

As such, for DM-photon scattering with a cross section $\sigma \sim \lambda^4$, for example, constraints from structure formation will be much weaker at longer wavelengths. Indeed in this case DM which is consistent with the constraint from  ref.~\cite{Boehm:2014vja} can have a cross section as large as  $\sigma (\lambda = 10^{-4} \, \mathrm{m}) \lesssim 10^{-1} \cdot (m / \mathrm{GeV}) \cdot \sigma_T$  for observations of  $10^{-4} \, \mathrm{m}$ wavelength light.
Hence as we found in the previous section observations at longer wavelengths, in the infra-red and beyond, present the best prospect for observing DM-photon scattering.
For models where the cross section increases with $\lambda$ our method is complimentary to other direct constraints~\cite{Boehm:2014vja,Wilkinson:2013kia}, and presents the only way of accurately determining the wavelength dependence of the DM-photon cross section.
There are in principle further direct probes of DM-photon scattering, whose relative strengths depend on how $\sigma$ scales with wavelength $\lambda$ e.g. from the scattering of gamma-rays by DM in the halo from \textsc{Fermi}~\cite{Cirelli:2009dv,Ackermann:2012rg} or the observation of diffuse radio emission from DM scattering~\cite{Storm:2012ty} near strong radio sources.

Indirect constraints, such as on the charge of DM~\cite{CMBcharge,McDermott:2010pa} or interactions between DM particles themselves~\cite{McDermott:2010pa,Peter:2012jh,Kahlhoefer:2013dca}, will in principle be related to the DM-photon cross section. However the relative strength of constraints will depend on the particular model of DM e.g. hidden sector models~\cite{Foot:2004wz,Jaeckel:2013uva} or composite DM, such as dark atoms~\cite{Cline:2012is,Wallemacq:2013hsa} (or Standard Model composite states e.g.~\cite{Jacobs:2014yca}). As such constraints on the DM-photon cross section and other probes such as self-interactions or the charge of DM will give complimentary information.

\section{Conclusion \label{sec:conc}}
We have considered the observational consequences of Dark Matter particles, in the halo of spiral galaxies, scattering light emitted by luminous matter in the centre of the disc out to large radii. An illustration of this principle is shown in figure~\ref{fig:scattering_diag}. 
This is advantageous as a search strategy for DM, since the potential astrophysical backgrounds at large radii are smaller, as compared to searches for DM self-annihilation which focus on the  galactic centre region (e.g.~\cite{Daylan:2014rsa,Gordon:2013vta,Boehm:2014hva}).

Using measurements of light from M101 by the Dragonfly instrument~\cite{2014ApJ...782L..24V} (see  figure~\ref{fig:dm_photon_-28}) we showed that, for cross sections around $10^{-23} \cdot (m / \mathrm{GeV})$~cm$^2$, Dark Matter (with mass $m$) will scatter light out from the more luminous centre of the disc to its edge. This leads to an effective increase in the surface luminosity at large radii, where the emission from the disc itself is less intense. By marginalising over the emission from the stellar halo, whose radial profile is similar to that from DM, we set a $90\%$ upper limit on the cross section at this level.

We also considered the prospect for improving sensitivity to $\sigma_{\mathrm{DM}-\gamma}$ using multi-wavelength measurements of diffuse light. As shown in figure~\ref{fig:dm_photon_spectra} we found that the prospects for observation should be particularly good for models of DM with a cross section $\sigma_{\mathrm{DM}-\gamma}$ which increases with wavelength $\lambda$. This is because the contribution from prominent backgrounds such as Rayleigh scattering of dust~\cite{2012MNRAS.423L.117M,Menard21062010,2014ApJ...789..131H} and emission from the stellar halo~\cite{0004-637X-712-1-692} should be smaller at longer $\lambda$. Hence using spectral information would allow the signal from DM-photon scattering to be separated from potential backgrounds. Taking observations away from the disc where emission is expected to be lower would also be beneficial.

There is also the prospect of extending the search for DM-photon scattering to other astrophysical sites such as elliptical galaxies, clusters, Active Galactic Nuclei and Gamma-Ray Bursts~\cite{Fynbo:2014uda}. 
Note also that unlike decaying or self-annihilating DM, there would be no expected signal in objects without luminous matter in significant quantities, such as dwarf spheroidal galaxies~\cite{Abdo:2010ex,GeringerSameth:2011iw}.

We have presented a new way of probing the interactions of Dark Matter, complimentary to bounds on the DM-photon cross section from the CMB and large-scale structure~\cite{Boehm:2014vja,Wilkinson:2013kia,Buckley:2014hja} which are sensitive to different photon energies, as well as alternative search strategies such as indirect detection, direct detection and collider searches. 
The unique radius and wavelength-dependent profile of DM scattering is advantageous for setting stronger constraints in the future, though at present the variation of the backgrounds from the disc and stellar halo between galaxies is not well understood, making a statistical analysis difficult. 
However even a small amount of progress here, for example by rescaling the disk components and stacking the halo light for several galaxies at multiple wavelengths, would allow the DM- photon scattering cross section to be probed with more precision than has previously been possible.

\section*{Acknowledgements}
\vspace{-10pt}
The research of JHD and JS has been supported at IAP by  ERC project 267117 (DARK) hosted by Universit\'e Pierre et Marie Curie - Paris 6 and also for JS at JHU by National Science Foundation grant OIA-1124403 and for JHD at IPPP by the UK STFC. JHD thanks Ryan Wilkinson for helpful comments.

 \end{document}